\pdfoutput=1 
\documentclass[referee]{raa}            
\usepackage{graphicx,times}             
\usepackage{natbib}
\usepackage{amssymb,amsmath}
\bibpunct{(}{)}{;}{a}{}{,}

\def\apjs{{\it Astrophys.~J.~Suppl.}}
\def\apj{{\it Astrophys.~J.}}
\def\aj{{\it Astronom.~J.}}
\def\apjl{{\it Astrophys.~J.~Lett.}}

\def\mnras{{\it Mon.~Not.~R.~Astron.~Soc.}}

\def\aap{{\it Astron.~Astrophys.}}

\def \A&ARv{{\it Astron.~Astrophys.Rev.}}

\begin{document}

   \title{{Optical spectroscopy of classical Be stars in old open clusters}
}

   \volnopage{Vol.0 (20xx) No.0, 000--000}      
   \setcounter{page}{1}          

   \author{Madhu Kashyap Jagadeesh 
      \inst{1}
  \and Blesson Mathew
     \inst{1}
       \and K. T. Paul
       \inst{1}
    \and Gourav Banerjee
  \inst{1} 
      \and Suman Bhattacharyya
   \inst{1}
   \and R. Anusha
   \inst{2}
   \and Pramod Kumar S
      \inst{3}
   }

\institute{\textsuperscript{1}Department of Physics and Electronics, CHRIST (Deemed to be University), Hosur Main Road, Bengaluru, India; {\it madhu.kashyap@res.christuniversity.in}\\
\textsuperscript{2}Department of Physics and Astronomy, University of Western Ontario, London, ON N6A 3K7, Canada\\
\textsuperscript{3}Indian Institute of Astrophysics, Koramangala, Bengaluru, India\\
\vs\no
   {\small Received~~20xx month day; accepted~~20xx~~month day}}

\abstract{We performed the optical spectroscopy of 16 classical Be stars in 11 open clusters older than 100 Myr. Ours is the first spectroscopic study of classical Be stars in open clusters older than 100 Myr. We found that the H$\alpha$ emission strength of most of the stars is less than 40 \AA, in agreement with previous studies. Our analysis further suggests that one of the stars, [KW97] 35-12, might be a weak H$\alpha$ emitter in nature showing H$\alpha$ EW of -0.5 \AA. Interestingly, we also found that the newly detected classical Be star LS III +47 37b, might be a component of a possible visual binary system LS III +47 37, where the other companion is also a classical Be star. Hence, the present study indicates the possible detection of a binary Be system. Moreover, it is observed that all 16 stars exhibit lesser number of emission lines compared to classical Be stars younger than 100 Myr. Furthermore, the spectral type distribution analysis of B-type and classical Be stars for selected clusters points out that there might be a probability that the existence of classical Be stars can depend on the spectral type distribution of B-type stars present in the respective clusters.
\keywords{classical Be stars, optical spectroscopy, open clusters, spectral lines}
}

   \authorrunning{Madhu Kashyap Jagadeesh et al.}            
   \titlerunning{Spectroscopy of CBe stars in old open clusters}  

   \maketitle

%
%
\section{Introduction}           

Collins (1987) defined a classical Be (CBe) as ‘a non-supergiant B-type star whose spectrum has, or had for a while, one or more Balmer lines in emission'. They are main sequence stars having luminosity classes III--V. An in depth overview of CBe star studies carried out on different aspects is given in  Rivinius, Carciofi, \&
Martayan (2013) and Porter \& Rivinius (2003). Being a subset of massive B-type stars, these CBe stars are rapid rotators and are characterized by emission lines of different elements in their spectra (e.g. Banerjee
et al. 2021; Shokry et al. 2018; Aguayo et al. 2017; Paul et al. 2012; Mathew \& Subramaniam 2011), in addition to an infrared excess in the spectral energy distribution (Hartmann \& Cassinelli 1977; Gehrz et al.
1974). These emission lines and infrared excess originate from a surrounding circumstellar disc, which is an equatorial, gaseous, geometrically thin, decretion disc that orbits the central star  (Meilland et al. 2007).

Spectral analysis of different emission lines in CBe stars has become useful in better understanding the kinematics and geometry of the circumstellar disc and various other properties of the central star itself. Interestingly, the disc formation mechanism in CBe stars, which is known as the ‘Be phenomenon’, remains an open issue till date. Hence, several spectroscopic surveys are performed till date, both in optical (e.g.
Klement et al. 2019; Arcos et al. 2017; Koubsk et al. 2012; Dachs et al. 1986; Hanuschik 1986; Andrillat
\& Fehrenbach 1982) and in the near-infrared (e.g. Granada et al. 2011; Steele \& Clark 2001; Clark \& Steele 2000) wavelength bands for characterizing CBe star discs and better understanding the ‘Be phenomenon’. Focusing on similar aspects, numerous more spectroscopic surveys have also been done for CBe stars situated in different environments, such as fields (Banerjee et al. 2021), open clusters  (Mathew et al. 2008) and extragalactic regimes such as the Magellanic clouds  (Paul et al. 2012). 

Open clusters consist of stars having similar age, thus providing better opportunities to study CBe stars and their evolutionary stages. A few important works which studied CBe stars located in open clusters are by Martayan et al. (2010), Mathew et al. (2008) and McSwain \& Gies (2005). Using slitless spectroscopy, Mathew et al. (2008) detected 152 CBe stars in 42 young open clusters. The spectral lines identified in this study are described in Mathew \& Subramaniam (2011). Motivated by this study, we performed a similar type of spectroscopic survey to search for CBe stars in 71 open clusters older than 100 Myr Jagadeesh et al. (2021). Our study identified 13 CBe stars located in 11 old open clusters. Another two probable CBe stars, namely [KW97] 35-12 and HD 16080 were also observed. Moreover, we found that LS III +47 37 is a possible visual binary system containing a CBe star, LS III +47 37a (Monguio et al. 2017).

In the present paper, we performed the optical spectroscopy of these 15 CBe stars in 11 clusters. We also performed the optical spectroscopy of LS III +47 37b, the second star in the possible visual binary system LS III +47 37. Our present study confirms LS III +47 37b to be a CBe star for the first time. Hence, the optical spectroscopic study of a total 16 CBe stars is presented in this paper. Further in the article, we will use the expression “old open clusters” to refer
to the open clusters analyzed in the paper. The paper is organised as follows: Sect. 2 describes the observation techniques used for the study. In Sect. 3, we present the major results and analysis done in this work. Prominent results obtained from the present study are summarised in Sect. 4.

\section{Observations}

We obtained the medium resolution spectra of 7 among 16 of our sample CBe stars during October and December, 2020 using the Optomechanics Research (OMR) spectrograph Prabhu et al. 1998, mounted on the 2.34–m Vainu Bappu Telescope (VBT), situated at the Vainu Bappu Observatory (VBO), Kavalur, Tamil Nadu, India. These stars were selected since they are bright enough for observing with the 2.34--m telescope. The photographic sensor used in VBT is an Andor CMOS high speed read out sensor, containing 1024 $\times$ 256 pixels. It possesses pixel size equal to 23 $\mu$m. We obtained the spectra for 7 CBe stars in the wavelength region of 5500$-$7500 \AA, at settings centered at H$\alpha$ line, producing a resolving power of 1000.

The spectra for the remaining 9 CBe stars are obtained with the HFOSC instrument mounted on the 2.1--m Himalayan Chandra Telescope (HCT) located at Hanle, Ladakh, India. We observed these 9 stars during June 2021, based on the observation visibility of HCT. The spectral coverage is from 5500 -- 9000 \AA. The spectrum in this `red region' is taken with Grism 8 (5500 $-$ 9000 \AA), which in combination with 167l slit provides an effective resolution of 7 \AA~at H$\alpha$ ((Mathew \& Subramaniam 2011)). 

During both sets of observations, dome flats were taken along with halogen lamps. These were used for the purpose of flat fielding the images. Standard IRAF tasks were used for bias subtraction, flat field correction and spectral extraction. FeNe lamp spectra were obtained along with the object spectra for performing wavelength calibration. Lastly, the extracted raw spectra were wavelength calibrated and continuum normalized taking the help of IRAF tasks. The representative spectra for two of our sample CBe stars, HD 280460 and TYC 2679-432-1 observed using the VBT and HCT facilities, respectively. are shown in Fig. 1. The log of our observations is shown in Table 1.

\begin{figure}[h!]
\centering
\includegraphics[height=102mm,width=155mm]{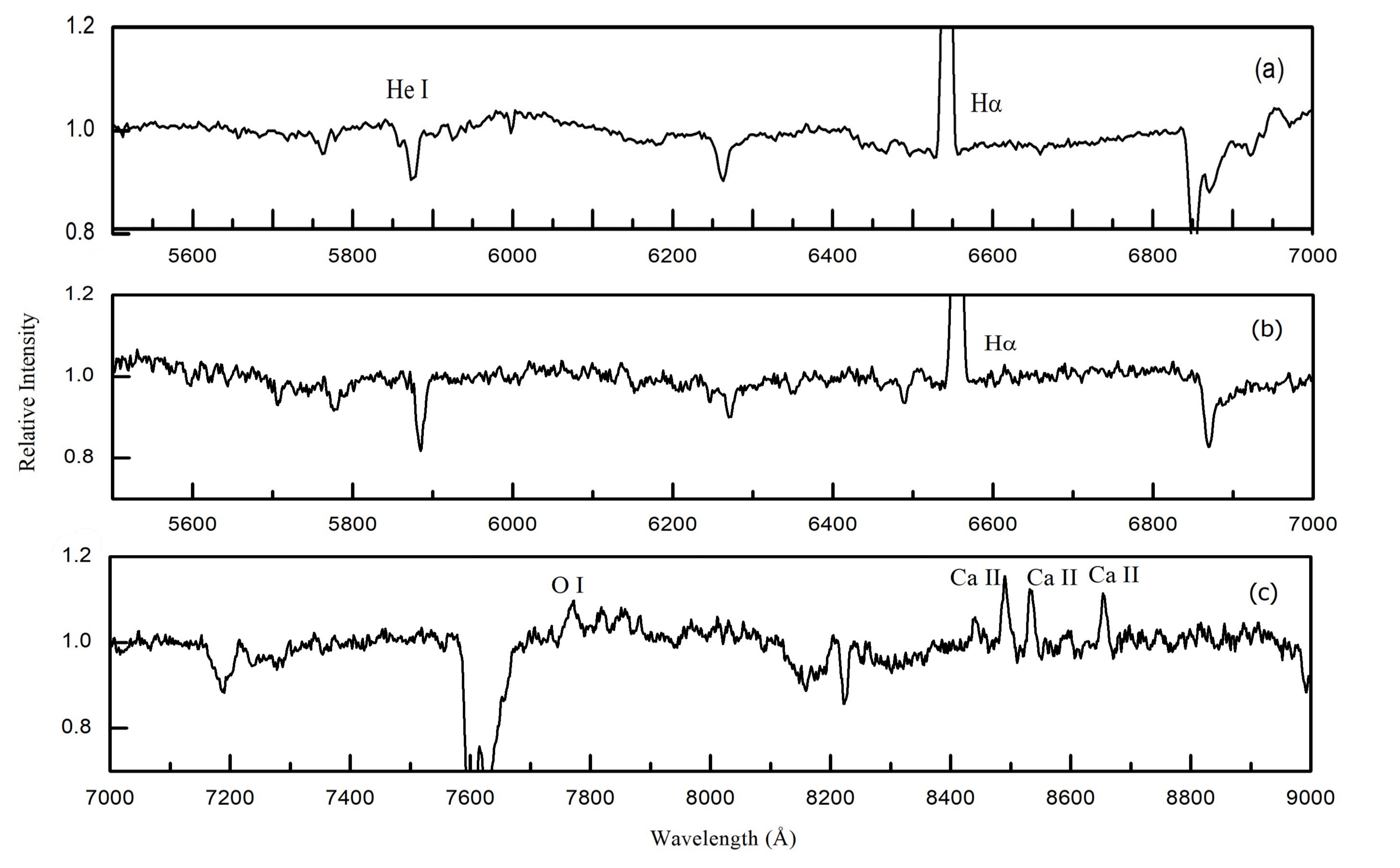} 
\caption{Representative spectra for two of our sample CBe stars, namely HD 280460 shown in panel a and TYC 2679-432-1  presented in panels b and c, respectively. While the spectrum of HD 280460 is obtained using the OMR instrument of the VBT facility, spectrum of TYC 2679-432-1 is taken with the HFOSC instrument mounted on the HCT facility.}
\label{fig1}
\end{figure}

\begin{table}[h!]
    \centering
    \caption{Observation log of 16 CBe stars identified in open clusters older than 100 Myr. The spectral types of 15 among 16 stars are obtained from Jagadeesh et al. (2021). For the star LS III +47 37b, we estimated the spectral type during the present study following the photometric technique described in Jagadeesh et al. (2021).}
\label{table1}
    \begin{tabular}{lllllllll}
    \hline
    VBT Observation \\ \hline
        Cluster \& Simbad ID \& Date of obs. \& RA \& DEC \& V \& Spectral \& Age \& Exposure \\ 
        ~ \& ~ \& (yyyy-mm-dd) \& (hh mm ss) \& (dd mm ss) \& (mag) \& type \& (Myr) \& time (s) \\ \hline
        Trumpler 2 \& HD 16080 \& 2020-12-20 \& 02 37 00 \& +55 54 41 \& 9.3 \& B2 \& 147 \& 1800 \\ 
        Tombaugh 5  \& SS 16 \& 2020-12-27 \& 03 47 50 \& +59 03 59 \& 11.4 \& B2 \& 199 \& 2400 \\ 
        NGC 1778 \& HD 280460 \& 2020-12-21 \& 05 08 13 \& +36 59 36 \& 9.7 \& B3 \& 158 \& 600 \\ 
        ~ \& HD 280461 \& 2020-12-21 \& 05 08 04 \& +36 58 27 \& 10.2 \& B6 \& 158 \& 1200 \\ 
        ~ \& HD 280462 \& 2020-12-21 \& 05 08 02 \& +37 03 03 \& 10.2 \& B6 \& 158 \& 1200 \\ 
        IC 2156  \& LS V +24 11 \& 2020-12-21 \& 06 05 02 \& +24 09 28 \& 11.7 \& B2 \& 251 \& 2400 \\ 
        NGC 6709 \& BD+10 3698 \& 2020-10-06 \& 18 51 32 \& +10 19 10 \& 9.7 \& B2.5 \& 158 \& 120 \\ \hline
        HCT Observation \\ \hline 
        Ruprecht 144 \& SS 398 \& 2021-06-26 \& 18 33 52 \& -11 24 31 \& 12.1 \& B8 \& 158 \& 720 \\ 
        Trumpler 34 \& GSC 5692-0543 \& 2021-06-26 \& 18 39 49 \& -08 25 40 \& 12.1 \& B6 \& 125 \& 720 \\ 
        NGC 6709 \& [KW97] 35-12 \& 2021-06-26 \& 18 51 10 \& +10 23 25 \& 10.9 \& B8 \& 158 \& 720 \\ 
        Berkeley 47 \& TYC 1605-346-1 \& 2021-06-26 \& 19 28 31 \& +17 22 22 \& 12.3 \& B2 \& 158 \& 600 \\ 
        Berkeley 50 \& TYC 2679-432-1 \& 2021-06-26 \& 20 10 05 \& +34 55 44 \& 11.2 \& B0.5 \& 251 \& 600 \\ 
        Berkeley 90 \& UCAC3 274-184438 \& 2021-06-26 \& 20 35 41 \& +46 46 49 \& 12.9 \& B2 \& 100 \& 720 \\ NGC 7067 \& LS III +47 37a \& 2021-06-26 \& 21 24 12 \& +48 00 39 \& 13.2 \& B1 \& 100 \& 600 \\ 
        ~ \& LS III +47 37b \& 2021-06-26 \& 21 24 12 \& +48:00:40 \& 11.3 \& B8 \& 100 \& 600 \\ 
        King 20 \& GGR 148 \& 2021-06-26 \& 23 33 03 \& +58 27 44 \& 12.1 \& B2.5 \& 199 \& 600 \\ \hline
    \end{tabular}
\end{table}

\section{Results}
In this section we present the results from the analysis of the spectra of 16 CBe stars. They belong to spectral types B0 -- B8 Jagadeesh et al. (2021). Out of 16 stars, 10 are similar to or earlier than the B3 spectral type, whereas others are later than B5. There is no star of B4 or B5 types. We were not able to re-estimate the spectral types of the sample stars using spectroscopy since we could not obtain the spectra in the blue region, centred at H$\beta$ line. Hence, we adopted the spectral type for every star as estimated by Jagadeesh et al. (2021), using the photometric spectral classification method (shown in Table 1 of Jagadeesh et al. 2021).

In Sect. 3.1, we show the spatial distribution of our sample 16 CBe stars in the Galaxy as compared to a set of 150 CBe stars located in open clusters younger than 100 Myr. Sect. 3.2 includes the description of the spectral features observed in each of our sample of 14 CBe stars. Sect. 3.3 presents the analysis of the possible visual binary star system LS III +47 37. In Sect. 3.4, the H$\alpha$ equivalent width (EW) distribution found in our sample is compared with that observed in CBe stars located in young clusters (younger than 100 Myr). Sect. 3.5 provides a comparative study of different spectral features observed in CBe stars in field and open clusters. The last section includes the comparative analysis of the spectral type distribution of the sample of  B-type stars with that of CBe stars, both in clusters below and above 100 Myr, respectively.

\subsection{Spatial distribution of CBe stars in the Galaxy}
The polar plot in Fig. 2 represents the spatial distribution of our sample of 16 CBe stars above 100 Myr and 150 CBe stars below 100 Myr (taken from Mathew et al. 2008). The distances used for this analysis are taken from the Gaia DR3 estimates (Bailer-Jones 2022). The center point (or origin) of the plot is considered as the position of the Sun. The black and red dots represent CBe stars below 100 Myr and CBe stars above 100 Myr, respectively. Each concentric circle in the plot represents a distance range of 500 pc away from the Sun. We therefore conclude that the lack of distribution of CBe stars belonging to old open clusters in the 225$^{\circ}$ region, may be due to the smaller sample size of our study. This means that none of the CBe stars above 100 Myr are located in the region from the right side of the Perseus arm to the edge of Scutum Centaurus arm of the Galaxy. However, the large survey of young CBe stars in clusters below 100 Myr done by Mathew et al. (2008) shows a larger global coverage of distribution, when compared to our sample. 

\begin{figure}[h!]
\centering
\includegraphics[height=120mm,width=95mm]{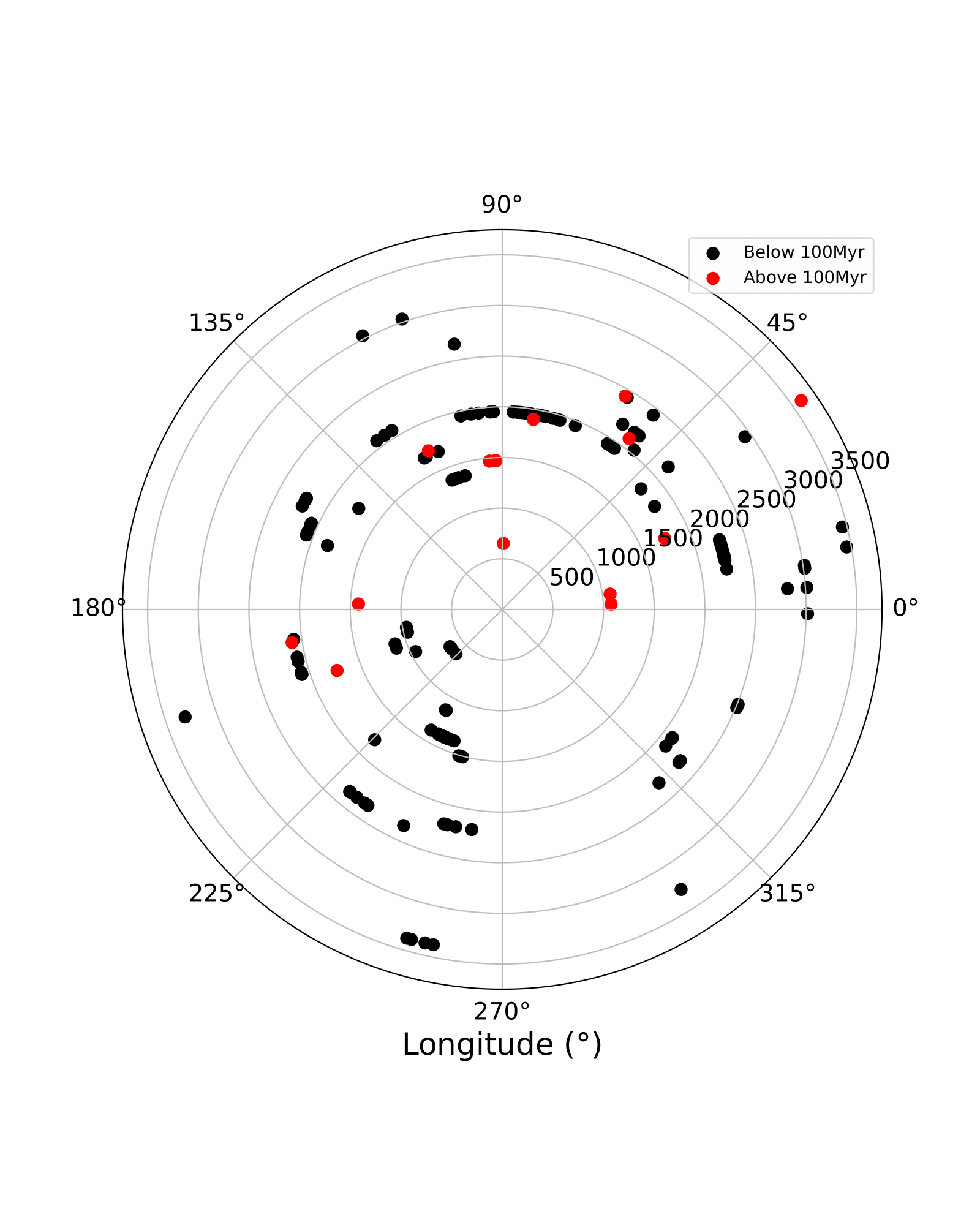} 
\caption{Spatial distribution of our sample of 16 CBe stars above 100 Myr and 150 CBe stars below 100 Myr (taken from Mathew et al. 2008) with respect to estimated distances from the Gaia DR3 data. It is noticed that all of our sample of 16 CBe stars in older clusters is located in the longitude range between 0 -- 225$^{\circ}$.}
\label{fig2}
\end{figure}

\subsection{Observed spectral features in individual stars}
The representative spectra for two of our sample CBe stars, HD 280460 and TYC 2679-432-1 is already shown in Fig. 1. In Table 2 of this section we describe the spectral features observed in every individual star. The relevant detailed literature for each star is mentioned in Jagadeesh et al. (2021).

\begin{table}[h!]
\centering
\caption{Spectral features observed in our sample of 16 CBe stars. The measured and corrected H$\alpha$ equivalent width (EW) values for every star are listed in columns 2 and 3, respectively. We estimated the corrected H$\alpha$ EW for all 16 stars following the method described in Sect. 3.4.1 of Banerjee et al. (2021). The (-) sign denotes emission, whereas positive value denotes absorption. The terms e, a and {\it eia} stands for ‘emission', ‘absorption' and ‘emission in absorption', respectively. }
\label{table2}
\begin{tabular}{llll}
\hline
VBT Observations \\ \hline
 Simbad ID  \& H$\alpha$ EW\textunderscore{m} \& H$\alpha$ EW\textunderscore{c}  \& Other features             \\ \hline
 
SS 16   \&-16.7 \AA \& -21.4 \AA 
\& Fe{\sc ii} 4926 \AA~(a) \\

HD 280460 \& -14.8 \AA \& -20.4 \AA
\& None  \\

HD 280461 \&-13.1 \AA \&-20 \AA
\& None   \\

HD 280462 \& -1.7 \AA \& -8.6 \AA
\& None     \\

LS V +24 11 \& -29.1 \AA  \&-33.8 \AA
\& None      \\

BD+10 3698 \& -5.8 \AA~(eia) \&-10.5 \AA
\& None     \\
HD 16080 \& -4.2 \AA~(eia) \& -8.9 \AA
\&  None     \\
 \hline
 
HCT Observations \\ 
\hline 
SS 398   \&  -19.4 \AA \& -27.3 \AA
\& P11 -- P13 (e) \\

GSC 5692-0543 \& -37.5 \AA \& -44.4 \AA
\& P11 -- P17 (e), O{\sc i} 7772 \AA~(a)\\

TYC 1605-346-1  \& -12.2 \AA \&-16.9 \AA
\& P11 -- P20 (e), He{\sc i} 5876 (a), O{\sc i} 7772 \AA~(a)\\

TYC 2679-432-1 \& -12.1 \AA \& -16 \AA 
\&  Ca{\sc ii} triplet 8498, 8542, 8662 \AA~(e), He{\sc i} 5876 \AA~(a)\\

UCAC3 274-184438 \& -32.1 \AA \& -36.8 \AA
\& He{\sc i} 5876 \AA~(a) \\

LS III +47 37a \& -2.1 \AA \& -6.0 \AA
\& He{\sc i} 5876 \AA~(a) \\

LS III +47 37b \& -4.2 \AA \& -12.1 \AA \& O{\sc i} 7772 \AA~(e), He{\sc i} 5876 \AA~(a) \\

GGR 148  \& -13.7 \AA \& -18.4 \AA
\& He{\sc i} 5876 and Si{\sc ii} 6371 \AA~(a)  \\

$[KW97]$ 35-12 \& 7.4 \AA \& -0.5 \AA
\& He{\sc i} 5876 and O{\sc i} 7772 \AA~(a) \\ \hline
\end{tabular}
\end{table}

During our previous study Jagadeesh et al. (2021), we found two stars ([KW97] 35-12 and HD 16080) exhibiting H$\alpha$ in absorption. Both were confirmed to be B-type stars by Jagadeesh et al. (2021). Interestingly, H$\alpha$ in emission has been reported for both these stars in the past. This raised the possibility that they might be CBe stars which are passing through disc-less state. Showing variability in spectral line profiles is a common property of CBe stars. In extreme cases, a complete disappearance of the H$\alpha$ emission line takes place, indicating a disc-less phase in CBe stars. The spectrum then appears to be similar to that of a normal B-type star exhibiting photospheric absorption lines. Literature review shows that a good number of CBe stars have shown such a transient nature, at least once in their lifetime [a few recent works are][]{Marr et al. 2021; Cochetti et al. 2021; Ghoreyshi et al. 2021; Klement et al. 2019; Mathew et al. 2013;
Mathew \& Subramaniam 2011)}. Hence, we studied the two stars [KW97] 35-12 and HD 16080 to analyse the variability in their spectral features.

\subsubsection{[KW97] 35-12}
A less studied star, [KW97] 35-12 showed H$\alpha$ in absorption when observed by us on July 08, 2015 (Jagadeesh et al. 2021) using slitless spectroscopy. However, when observed with the HCT, we found that H$\alpha$ emission exists below the continuum for this star with an EW of -0.5 \AA. This can be suggestive of the fact that it might be a weak H$\alpha$ emitter in nature, which can be confirmed through further spectroscopic studies. It is to be noted that such weak H$\alpha$ emitters have been identified already in the case of field CBe stars (Banerjee et al. 2021). Other spectral features observed for this star are listed in Table 2.

\subsubsection{HD 16080}

Interestingly, H$\alpha$ was visible in absorption when we observed HD 16080 on January 02, 2016 (Jagadeesh et al. 2021) using slitless spectroscopy. In our present study, we found that H$\alpha$ exhibits emission in absorption profile having an EW of -8.9 \AA~when observed with the VBT facility. Apart from H$\alpha$, no other spectral feature is visible for HD 16080. 

Our results thus confirm that both these objects are CBe stars in nature. However, we could not confirm whether they have passed through some disc-less episodes in 2015 (in the case of [KW97] 35-12) and 2016 (for HD 16080), since we could not measure the H$\alpha$ EW during our previous study using slitless spectroscopy (Jagadeesh et al. 2021). Moreover, both these stars being less studied, no literature exists reporting H$\alpha$ in absorption using spectroscopic analysis till date.

\subsection{LS III +47 37: a possible visual binary system}
LS III +47 37 is a visual pair system located in the cluster NGC 7067 ((Mongui'o et al. 2017). Situated at a distance of 3600 pc, NGC 7067 is having an age of 100 Myr ((Becker 1965). LS III +47 37a, having Gaia ID 2164725643889832960, is one of the components of this system. (Mongui'o et al. 2017 detected it to be a CBe star of spectral type close to B0.5. When observed with HCT, we noticed that this star exhibits H$\alpha$ in emission having an EW of -6 \AA. He{\sc i} 5876 \AA~line is also visible in absorption. 

LS III +47 37b is the other component of the system. We estimated the spectral type of this star to be B8 using the photometric technique mentioned in Jagadeesh et al. (2021). The V magnitude is obtained from Monguio et al. (2017). Interestingly, H$\alpha$ showed single peak emission with EW of -4.2 \AA~(without absorption correction), when we observed it with HCT. Our study thus confirms this object to be a CBe star for the first time. Additionally, emission line of O{\sc i} 7772 \AA~is detected in case of LS III +47 37b. Prominent absorption line of He{\sc i} 5876 \AA~is also visible.

Moreover, since the other component (LS III +47 37a) of this system was also identified as a CBe star, this might become an interesting pair of CBe stars, if it is confirmed to be a binary system. Looking into the newly available data from Gaia DR3, we found that the distance of LS III +47 37a is reported as 3412 pc. For the other star LS III +47 37b no distance is available in Gaia DR3. So we could not confirm the binary nature of this system. Binary systems containing one CBe star and another compact object companion has been detected in several previous studies [e.g.][]{. Bhattacharyya et al. 2022; Kennea et al. 2021; Okazaki
\& Negueruela 2001}. However, literature review reveals a scarcity of studies related to binary systems containing both CBe stars. Hence, we suggest further investigation of the LS III +47 37 system to confirm its binary nature.

\subsection{H$\alpha$ equivalent width distribution}
Before looking into other spectral lines, we first checked the H$\alpha$ EW distribution in our sample. Our results show that the H$\alpha$ EW for all 16 stars range within -0.5 to -44.4 \AA. Out of 16, 15 ($\sim$ 94\%) stars show H$\alpha$ EW $<$ -40 \AA. This result is in agreement with Banerjee et al. (2021) who found that the H$\alpha$ EW values in CBe stars are mostly lower than -40 \AA. The star GSC 5692-0543 among our sample exhibits the highest H$\alpha$ EW, -44.4 \AA, and is the only star possessing H$\alpha$ EW more than -40 \AA. Also, Mathew \& Subramaniam (2011) noticed that around 80\% of their sample of young cluster CBe stars have H$\alpha$ EW between -1 to -40 \AA.

The H$\alpha$ EW distribution for our sample of CBe stars is shown in Fig. 3. Here, the black color bars represent CBe stars in clusters below 100 Myr obtained from Mathew \& Subramaniam (2011) and the red color bars represent our sample CBe stars in clusters above 100 Myr. It is noticed from the figure that 5 out of our sample of 16 stars fall in the bin range of -10 to -20 \AA, whereas another 3 are located between -20 to -30 \AA~bin range. A similar peak is noticed in the sample of Mathew \& Subramaniam (2011). This result indicates that the H$\alpha$ EW for CBe stars in young or old clusters might not get affected due to different environments having different metallicity. However, while we observed that most of our sample stars fall in the bin range of -10 to -20 \AA, Mathew \& Subramaniam (2011) found most of their sample stars within -10 to -30 \AA~bin range. 

\begin{figure}[h!]
\centering
\includegraphics[height=70mm,width=90mm]{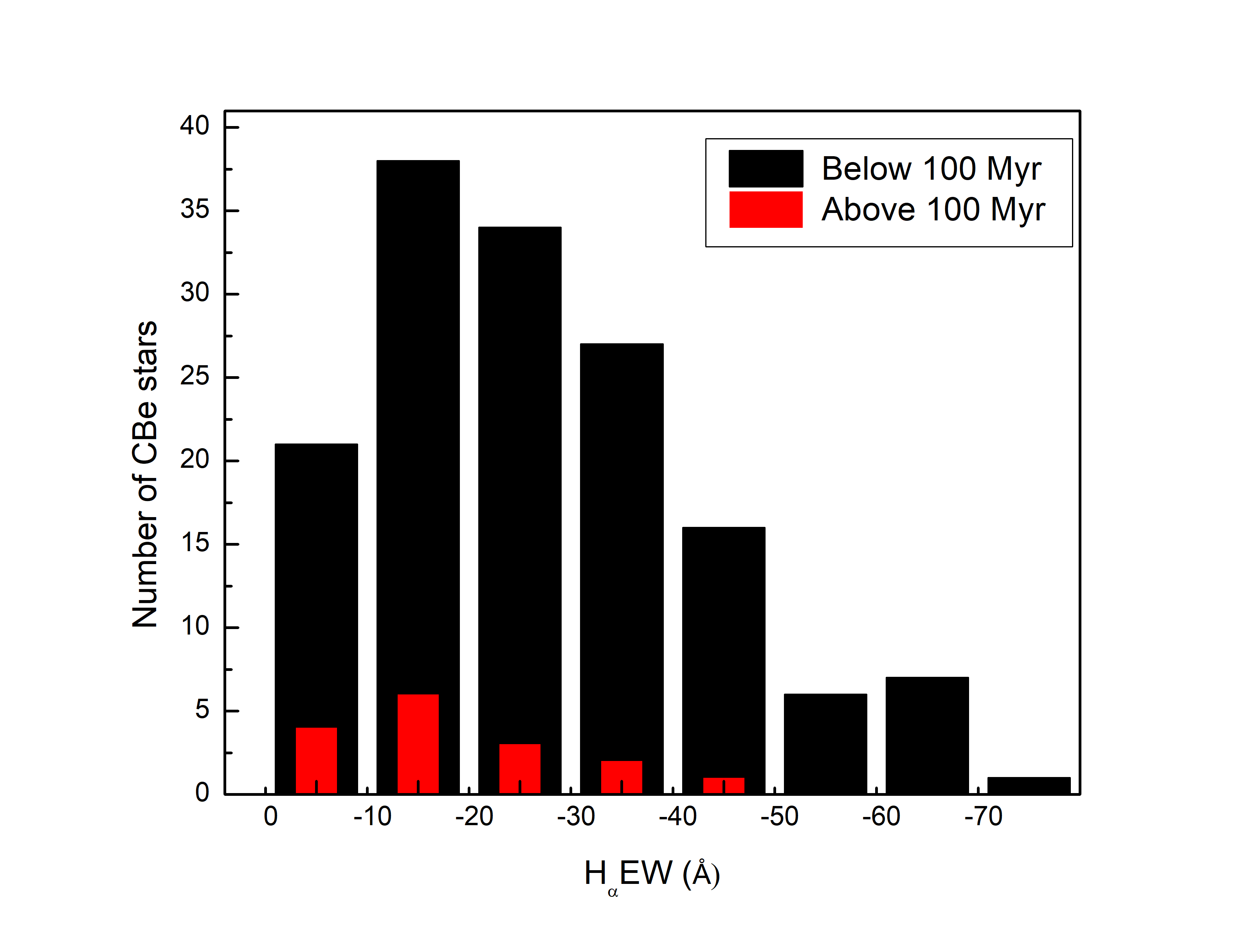} 
\caption{H$\alpha$ EW distribution for our sample of CBe stars. The black color bars represent CBe stars in clusters below 100 Myr obtained from Mathew \& Subramaniam (2011) and the red bars represent our sample CBe stars in clusters above 100 Myr.}
\label{fig3}
\end{figure}

Moreover, we performed a comparative analysis of the distribution of H$\alpha$ EW against the spectral types for CBe stars in clusters below and above 100 Myr. Fig. 4 presents this plot where the black dots represent CBe stars below 100 Myr and red dots represent CBe stars above 100 Myr, respectively. It is observed from the figure that the H$\alpha$ EW for early type CBe stars (within B1 -- B2) situated in young clusters span almost the whole range from -0.5 to -70 \AA. The H$\alpha$ EW range appears to decrease as we move to late types. Any such trend could not be confirmed in case of CBe stars in clusters above 100 Myr. However, it seems that for CBe stars above 100 Myr, H$\alpha$ EW range is comparatively lower than those with age below 100 Myr.

\begin{figure}[h!]
\centering
\includegraphics[height=80mm,width=100mm]{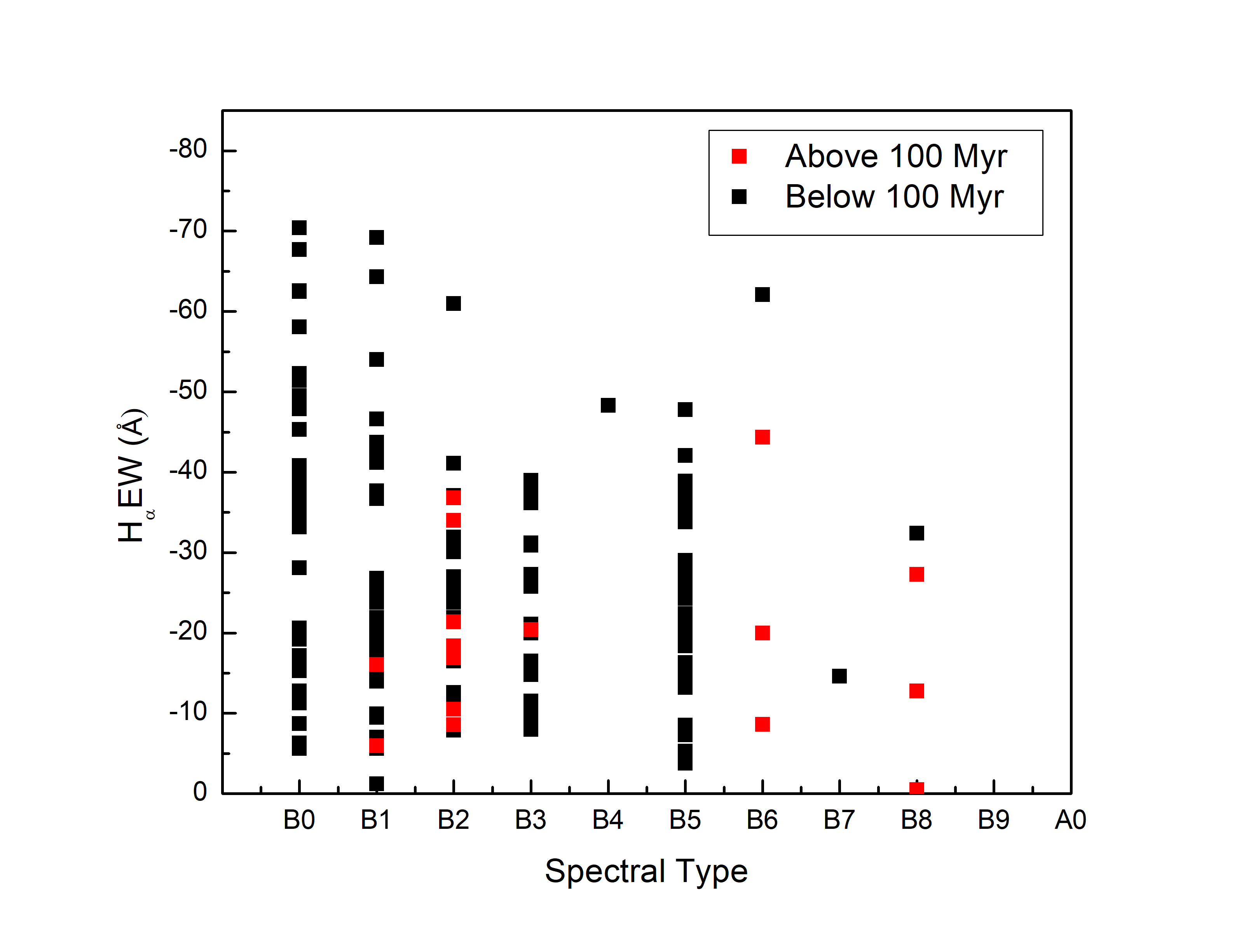} 
\caption{Distribution of H$\alpha$ EW against the spectral types for CBe stars in clusters below and above 100 Myr. Here, the black dots represent CBe stars below 100 Myr and red dots represent CBe stars above 100 Myr, respectively.}
\label{fig4}
\end{figure}

\subsection{Spectral features observed in CBe stars in young and old clusters}
Among our sample of 16 CBe stars, we found that the H$\alpha$ line is visible as a single-peaked emission profile for 12 stars. H$\alpha$ emission is detected to exist below the continuum for only one star, [KW97]35--12 (discussed in Sect. 3.2). The rest two stars (BD+10 3698 and HD 16080) show H$\alpha$ line in emission in absorption profile. 
It is also observed that 6 among the 15 stars exhibit He{\sc i} 6678 \AA~line in absorption. Moreover, Fe{\sc ii} 4926 \AA~absorption line is seen in case of only one star, i.e. SS 16.

We detected emission lines belonging to the hydrogen Paschen series for two stars, i.e., GSC 5692-0543 and TYC 1605-346-1, which showed the series lines from P11 - P17 and P11 - P20, respectively. One other star, SS 398 shows Paschen lines (P11-P13) in absorption. Furthermore, out of these 8, only 3 stars (GSC 5692-0543, TYC 1605-346-1 and [KW97]35--12) are seen to exhibit O{\sc i} 7772 \AA~absorption line. On the contrary, O{\sc i} 7772 \AA~line is seen in emission only in one case, for LS III+47 37b. Interestingly, Ca{\sc ii} triplet emission lines are also found in one among these 8 stars, TYC 2679-432-1. This result indicates that Ca{\sc ii} emission lines are visible in $\sim$ 13\% of our sample stars, in good agreement with Banerjee et al. (2021). Circumbinary discs have been suggested as one of the probable Ca{\sc ii} triplet line forming region in CBe stars by Banerjee et al. (2021). However, TYC 2679-432-1 being an unstudied star, we could not confirm whether it belongs to a binary system.

Surprisingly, we did not observe any emission line of Fe{\sc ii} or even the O{\sc i} 8446 \AA~emission line in any of our sample of CBe stars. Both these are commonly observed emission features in CBe stars, whether in young clusters Mathew \& Subramaniam 2011 or in the field population Banerjee et al. 2021. Moreover, although rare, no emission features of Mg{\sc ii} or Na{\sc i} were identified in the spectra of our sample. From the present study, we observed CBe stars older than 100 Myr exhibit lesser number of emission features compared to CBe stars younger than 100 Myr.

\subsection{Distribution of B-type and CBe stars in selected open clusters below and above 100 Myr}
For checking the incidence of CBe stars with respect to B-types in our studied clusters, we performed individual distribution study of B-type and CBe stars for all 11 clusters older than 100 Myr. The distribution plots of B-type and CBe stars located in all these 11 clusters is shown in Fig. 5. Here, the blue bars represent the number of B-type stars in respective clusters obtained from the literature. The orange bars represent the CBe stars identified by Jagadeesh et al. (2021) situated in the respective clusters.

\begin{figure}
\centering
\includegraphics[height=110mm, width=150mm]{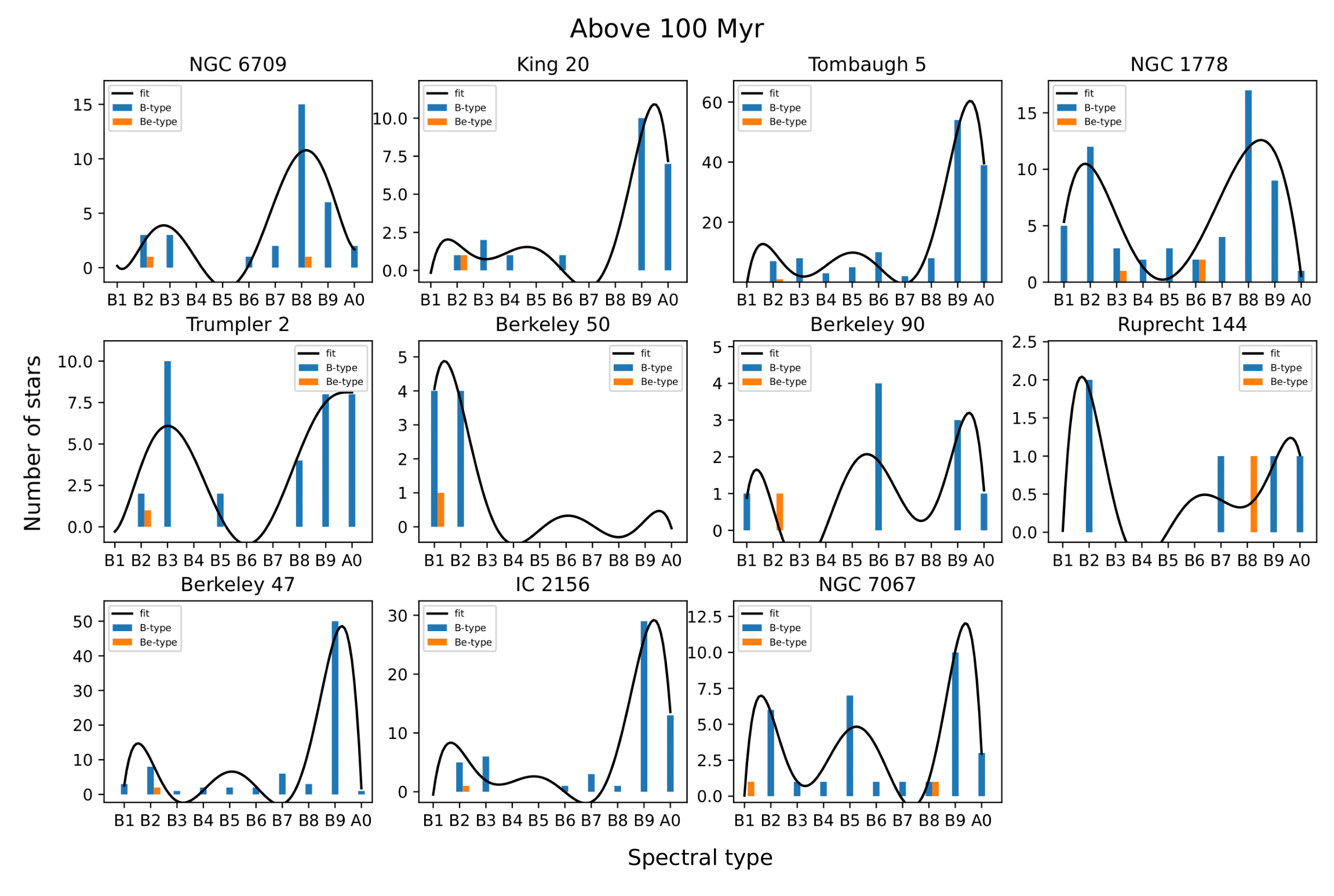} 
\caption{Plots for all 11 old open clusters showing the distribution of B-type and CBe stars in respective clusters. Here, the blue bars represent the number of B-type stars in respective clusters obtained from the literature.}
\label{fig5}
\end{figure}

It is visible from Fig. 5 that the B-type stars show a peak incidence at B8 – B9, in the case of 6 clusters. A bimodal distribution in spectral types is noticed for B-type stars in the clusters NGC 1778 and Trumpler 2, peaking at B1 – B2 and again at B8-B9. However, any such incidence could not be firmly identified for the rest 3 clusters (Berkley 50, Berkley 90 and Ruprecht 144). Interestingly, it is visible that the CBe stars identified by Jagadeesh et al. (2021) are also located around these peaks where B-type stars are situated, except for one case, NGC 1778, where the identified CBe stars are found to be of B6 type. It is interesting to notice that CBe stars also seem to follow such a trend similar to B-types in respective clusters.

So for further checking, we also constructed the distribution plots for CBe stars with respect to B-types in young open clusters (shown in Fig. 6). The selected 10 young clusters containing CBe stars are taken from Mathew et al. (2008). We have not included any rich open cluster for the analysis since the distribution is found to preferentially peak at early B-type (see the discussion in Mathew et al. 2008). In Fig. 6 the blue and orange bars represent B-type and CBe stars in their parent clusters, respectively. It is seen that B-type stars show a peak incidence near late types, i.e. B8-A0, in the case of 4 clusters, respectively. A bimodal distribution in spectral types is noticed for B-type stars for the rest 6 clusters. While in clusters Berkley 86, NGC 1624, NGC 7235, NGC 637 and NGC 7128, the peaking is observed near B2 and B8 -- A0 region, whereas the remaining cluster Bochum 2 exhibits bimodal peaks near B5 -- B6 and B8 -- B9, respectively. Similar to the case of clusters above 100 Myr (Fig. 5), it is found here also that the CBe stars identified by Mathew et al. (2008) are mostly located around these peaks where B-type stars are situated, except for 2 cases, NGC 146 and NGC 637 where the identified CBe stars are found to be of B6 type. Our result thus suggests that there might be a probability that the existence of CBe stars can depend on the spectral type distribution of B-type stars.

\begin{figure}[h!]
\centering
\includegraphics[height=110mm, width=150mm]{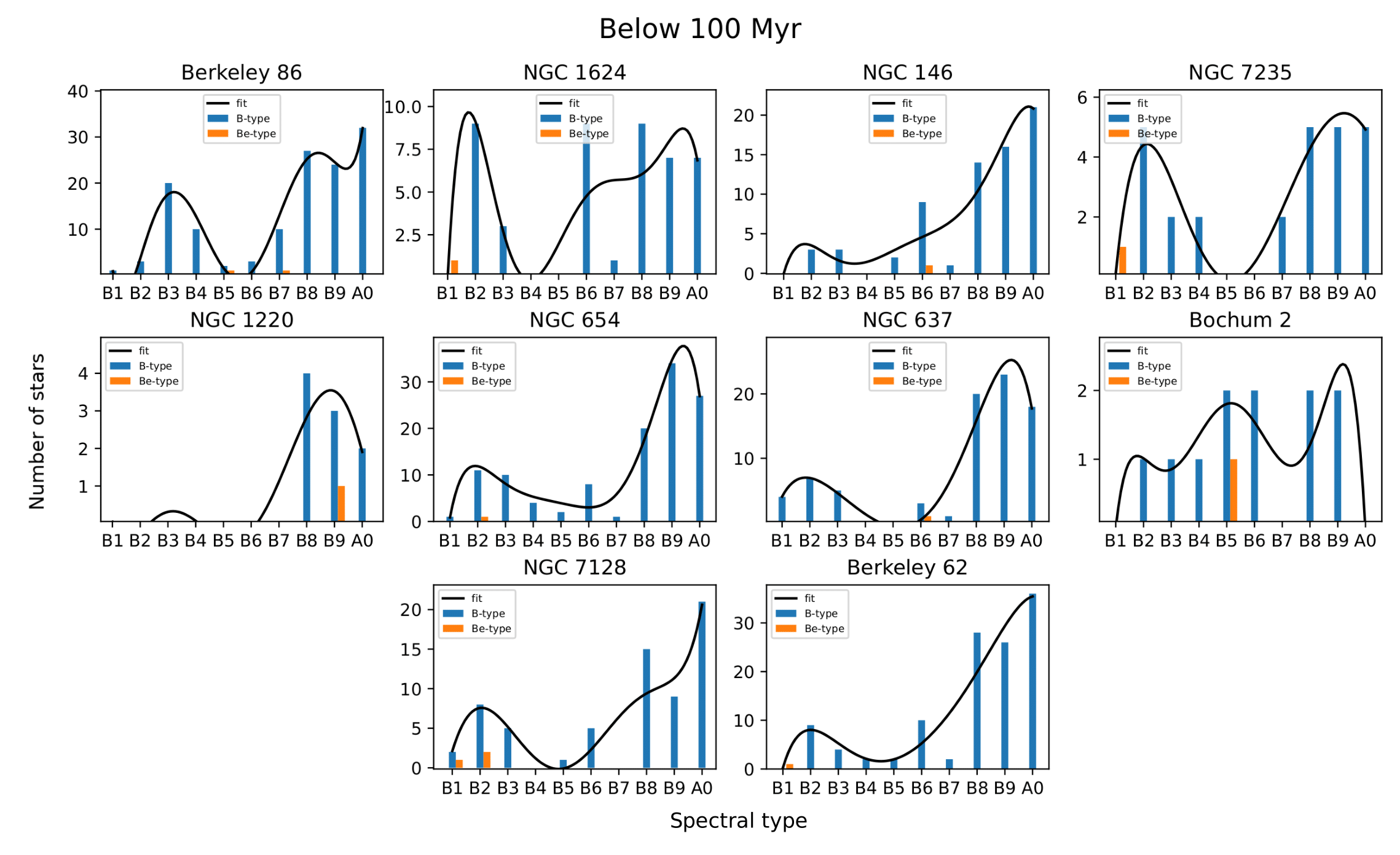} 
\caption{Distribution plots for CBe stars with respect to B-types in 10 selected clusters (taken from Mathew et al. 2008) younger than 100 Myr. Here, the blue and orange bars represent B-type and CBe stars in their parent clusters, respectively.}
\label{fig6}
\end{figure}


\section{Conclusions}
As a follow-up study with that of Jagadeesh et al. (2021), we performed the optical spectroscopy of a sample of 16 CBe stars in 11 old open clusters in the present paper. Ours is the first spectroscopic study of CBe stars in clusters older than 100 Myr. The prominent results obtained from our study are summarized below:

\begin{itemize}
\item Our study confirms that all 16 stars of our sample are CBe stars. One among these 16 stars, namely LS III +47 37b is a new detection having spectral type of B8. It belongs to the possible visual binary system LS III +47 37, where the other companion is also a CBe star as confirmed by the present study.

\item We found that these 16 stars exhibit lesser number of emission features compared to CBe stars which are younger than 100 Myr Mathew \& Subramaniam 2011. None of them show any Fe{\sc ii} emission lines. Interestingly, O{\sc i} 8446 \AA~emission feature is noticed in only one star (LS III +47 37b) and Paschen lines are found to be present in emission in only two cases, i.e., SS 398 and GSC 5692-0543. However, these are commonly observed features in the optical spectra of CBe stars, whether in fields  Banerjee et al. 2021 or in young clusters Mathew \& Subramaniam 2011. Hence, our study found that CBe stars belonging to clusters older than 100 Myr exhibit lesser number of emission features compared to CBe stars which are younger than 100 Myr, which is what is usually expected.

\item Our analysis also suggests that one out of these 16 stars, [KW97] 35-12, might be a weak H$\alpha$ emitter in nature showing H$\alpha$ EW of -0.5 \AA.

\item Moreover, we detected that the H$\alpha$ EW for 15 ($\sim$ 94\%) out of 16 of our sample stars is $<$ -40 \AA, in agreement with Banerjee et al. (2021). Interestingly, our comparative analysis with CBe stars from young clusters (younger than 100 Myr) points out that the H$\alpha$ EW for CBe stars in young or old clusters might not get affected due to different environments having different metallicity.


\item Furthermore, we performed individual distribution study of B-type and CBe stars for all clusters. Our analysis points out that there might be a probability that the existence of CBe stars can depend on the spectral type distribution of B-type stars present in the respective clusters.


\end{itemize}

Apart from Jagadeesh et al. (2021), quite a number of studies by several authors have identified different types of emission-line stars in the recent years (e.g. Li 2021; Anusha et al. 2021; Bhattacharyya et al. 2021;
Shridharan et al. 2021; Wang et al. 2022). Ours is thus a timely study regarding CBe star research. This study will help in motivating the entire Be star community about the need to further identify and study CBe stars in other older clusters. Detection and study of more CBe stars in fields and both younger and older clusters may provide new insights about the `Be pheomenon' in diverse environments. Moreover, effects of metallicity in CBe stars has already been noticed in several previous studies  (e.g. Maeder et al. 1999;
Keller 2004; Martayan et al. 2006, 2007a,b). Hence, further investigations of CBe stars in diverse metallicity environments (such as clusters and fields) might also provide clues about the mechanisms of disc formation and dissipation.

\section*{Acknowledgements}

We would like to thank the staff of VBO who took the observations using the Vainu Bappu Telescope (VBT) at Kavalur, India. We also thank our colleague Dr. Arun Roy for providing help during the observations. This research has made use of the WEBDA data base, maintained by the Institute for Astronomy of the University of Vienna. Moreover, we have used the SIMBAD data base, operated at CDS, Strasbourg, France. We thank the SIMBAD database for providing necessary help. Lastly, we would like to acknowledge the VizieR online catalog, which has been available since 1996 and was described in a paper published in A\&AS, 2000, 143, 23. 




\bibliographystyle{mnras}
\bibliography{reference}





\appendix



\label{lastpage}

\end{document}